%% file: Jhat_w0wa.tex
\documentclass[
reprint,
prd,
amsmath,amssymb,aps,
nofootinbib
]{revtex4-2}
\usepackage [latin1]{inputenc}
\usepackage{graphicx}
\usepackage{dcolumn}
\usepackage{bm}
\usepackage{amsmath}
\usepackage[dvipsnames]{xcolor}

\input{camille_setup}

\definecolor{mygreen}{RGB}{0,180,60}

\newcommand{\hJ}{\hat{J}}

\newcommand{\HH}{\mathcal{H}}

\usepackage{hyperref}

\begin{document}

\title{The impact of evolving dark energy on the Weyl potential measured from the Dark Energy Survey Year 3 data}

\author{Benedetta Rosatello}
\email{benedetta.rosatello@unige.ch}
\author{Gen Ye}
\email{gen.ye@unige.ch}
\author{Maria Berti}
\email{maria.berti@unige.ch}
\affiliation{%
D\'epartement de Physique Th\'eorique and Center for Astroparticle Physics, Universit\'e de Gen\`eve, Quai E. Ansermet 24, CH-1211 Geneva 4, Switzerland}%
\author{Isaac Tutusaus}
\email{tutusaus@ice.csic.es}
\affiliation{Institute of Space Sciences (ICE, CSIC), Campus UAB, Carrer de Can Magrans, s/n, 08193 Barcelona, Spain}
\affiliation{Institut d'Estudis Espacials de Catalunya (IEEC), Edifici RDIT, Campus UPC, 08860 Castelldefels (Barcelona), Spain}
\affiliation{Univ Toulouse, CNES, CNRS, IRAP, 14 Av.~Edouard Belin, 31400 Toulouse, France}
\author{Nastassia Grimm}
\email{nastassia.grimm@physics.ox.ac.uk}
\affiliation{Department of Physics, University of Oxford, Denys Wilkinson Building, Keble Road, Oxford OX1 3RH, United Kingdom}
\affiliation{Institute of Cosmology \& Gravitation, University of Portsmouth, Portsmouth, PO1 3FX, United Kingdom}
\affiliation{%
D\'epartement de Physique Th\'eorique and Center for Astroparticle Physics, Universit\'e de Gen\`eve, Quai E. Ansermet 24, CH-1211 Geneva 4, Switzerland}
\author{Camille Bonvin}
\email{camille.bonvin@unige.ch}
\affiliation{%
D\'epartement de Physique Th\'eorique and Center for Astroparticle Physics, Universit\'e de Gen\`eve, Quai E. Ansermet 24, CH-1211~Geneva 4, Switzerland}%

\date{\today}

\begin{abstract}
Measurements from the Dark Energy Survey (DES) Year 3 data have shown that the Weyl potential -- the sum of the spatial and temporal distortions of the geometry -- evolves more slowly than predicted by General Relativity, assuming a $\Lambda$CDM background evolution. An evolving dark energy with a phantom crossing, as preferred by the Dark Energy Spectroscopic Instrument (DESI), is expected to decrease the depth of the gravitational potentials through a stronger acceleration than in $\Lambda$CDM, potentially solving the tension with General Relativity. In this paper, we show that $w_0w_a$CDM models indeed reduce the tension with respect to $\Lambda$CDM, down to a level of $1.6-2.2\sigma$, depending on the treatment of CMB lensing. This reduction is not due to an increase in the Weyl potential's uncertainties, but truly to the impact of the evolving background on the theoretical predictions in General Relativity. More data are needed to robustly determine if evolving dark energy fully explains the low value of the Weyl potential at intermediate redshifts, or if modifications of gravity or interactions in the dark sector are needed, which could simultaneously stabilize the phantom crossing indicated by DESI.
\end{abstract}

\maketitle

\section{Introduction}

One of the main goals of cosmological observations is to determine the key ingredients governing our Universe, i.e.\ to understand how gravity works at cosmological scales, and to uncover the nature of dark energy and of dark matter. Information on these ingredients can be obtained in two ways. The first one consists in looking at the background evolution of the Universe, more precisely at its expansion rate, by measuring distances and redshifts. The second one is to measure the rate at which perturbations -- in the matter content and in the geometry -- grow with time. In a given theory of gravity, the evolution of the background and that of the perturbations are directly linked. In other words, the expansion rate of the Universe predictably affects the growth of perturbations.

Until recently, the measured expansion rate was in excellent agreement with the prediction of the $\Lambda$CDM model~\cite{Planck:2018vyg}. In that context, testing the laws of gravity can be done by measuring the evolution of perturbations, assuming a background evolution consistent with an equation of state $w=-1$, and determining whether the measured perturbations agree with the predictions of General Relativity (GR) or any other theory of interest that reproduces an effective $w=-1$~\cite{Clifton:2011jh,Koyama:2015vza,Joyce:2016vqv}. The outcome of those tests showed that galaxy velocities measured from redshift-space distortions are in good agreement with the GR predictions over a wide redshift range (see, e.g., Fig.~1 in~\cite{Grimm:2024fui}). In contrast, the Weyl potential -- the sum of the spatial and temporal distortions of the geometry -- measured from gravitational lensing, exhibits a mild tension at the level of around 3$\sigma$  with respect to the GR predictions~\cite{Tutusaus:2023aux}. More precisely, the Weyl potential is too low compared to the GR prediction in the first two bins of the Dark Energy Survey\footnote{\url{https://www.darkenergysurvey.org}} (DES)\, Year 3 (Y3) lens sample\,\footnote{A similar analysis has also been performed recently in~\cite{Zhang:2026gcg}, using weak lensing catalogs from the Kilo-Degree Survey and lens samples from the Baryon Oscillation Spectroscopic Survey and the 2-degree Field Lensing Survey, reporting larger error bars than the DES~Y3 analysis in~\cite{Tutusaus:2023aux} and finding no significant deviations from GR.}.  
This low value of the Weyl potential at low redshifts can be interpreted as a $\sigma_8$ tension~\cite{Abdalla:2022yfr}.

While the analysis in~\cite{Tutusaus:2023aux} was based on the assumption of a $\Lambda$CDM background evolution, the situation has recently changed with the advent of novel distance measurements from the Dark Energy Spectroscopic Instrument (DESI)\,\footnote{\url{https://www.desi.lbl.gov}}. Combined with luminosity distance measurements from type Ia supernovae (SNe) and with Cosmic Microwave Background (CMB) measurements, these results point to an evolving dark energy with a time-dependent equation of state that differs from $w = -1$~\cite{DESI:2024mwx,DESI:2025zgx}. This time dependence directly influences the evolution of perturbations. For example, during the phase of stronger acceleration seen by DESI above $z\simeq 0.4$ with $w<-1$, the Weyl potential decays faster than in $\Lambda$CDM, while at later times, when $w>-1$, it decays more slowly. Various independent analyses have reconstructed the time evolution of dark energy, with diverse methods, consistently pointing towards a phantom crossing at low redshift~\cite{DESI:2024aqx,Ye:2024ywg,Berti:2025phi,Fazzari:2025lzd,Gonzalez-Fuentes:2025lei}. The goal of this paper is to determine whether the tension between the measured and predicted GR values of the Weyl potential can be resolved by such an evolving dark energy component with a phantom crossing. 

To answer this question, we measure the Weyl potential at the four redshifts of the DES Y3  \texttt{MagLim} lenses, 
following the method developed in~\cite{Tutusaus:2022cab,Tutusaus:2023aux} and allowing for an evolving dark energy component. We parameterize the latter through its equation of state using the Chevallier-Polarski-Linder (CPL) parametrization~\cite{Chevallier:2000qy, Linder:2002et}. We then constrain the dark energy time evolution parameters $w_0$ and $w_a$, together with the Weyl potential, combining DES data with Planck, DESI Baryon Acoustic Oscillations (BAO), and supernovae luminosity distances. We find that the measured Weyl potential is almost insensitive to the background evolution, giving very similar results in $\Lambda$CDM and in $w_0w_a$CDM models. The theoretical prediction of the Weyl potential in GR is, however, 
sensitive to the choice of $w_0 $ and $w_a$.  We find that it is below the $\Lambda$CDM value at intermediate redshift, therefore reducing the tension with the measured Weyl potential. 

The size of the residual tension depends on how the lensing contribution to the CMB power spectra is modelled, and it ranges between 1.6--2.2$\sigma$ in the second redshift bin. In contrast, in $\Lambda$CDM the tension in that bin (which is the most discrepant one) ranges between 2.1--3.1$\sigma$.
This shows that evolving dark energy could be a possible explanation for the low values of the Weyl potential at intermediate redshift. More data are, however, needed to determine if it is truly enough to change the background evolution, or if one also needs to modify the evolution of perturbations directly, either by changing the laws of gravity or by including additional interactions in the dark matter sector. Interestingly, such modifications are required
in order to obtain a stable phantom crossing~\cite{Ye:2024ywg,Chudaykin:2024gol,Pan:2025psn,Wolf:2024stt,Wolf:2025acj,Wolf:2025jed,Wolf:2025jed,Akarsu:2024eoo}.

The rest of the paper is organized as follows: in Section~\ref{sec:formalism} we introduce the main observable used in our analysis, encoding the evolution of the Weyl potential. In Section~\ref{sec:data}, we discuss the set of data and our method to constrain the Weyl potential. The results are presented in Section~\ref{sec:results}, which shows the residual tension relative to the GR predictions. In Section~\ref{sec:discussion}, we explore models away from the output of our analysis, to further investigate the impact of $w_0w_a$CDM models on the Weyl potential. We conclude in Section~\ref{sec:conclusion}.

\section{Formalism}
\label{sec:formalism}

The main observable of the present work is the Weyl potential, which can be measured by combining galaxy-galaxy lensing with galaxy clustering. The key idea, presented in~\cite{Tutusaus:2022cab,Tutusaus:2023aux}, is to split the lensing and clustering angular power spectra into initial fluctuations at a chosen redshift $z_*$, wherein we assume that GR is recovered, and some free functions encoding the evolution of perturbations from $z_*$ to today. Instead of assuming a specific form for these functions, derived from a given theory of gravity, we reconstruct them directly from the data in a model-independent way.

We assume a perturbed Friedmann-Lema\^itre-Robertson-Walker universe, with a homogeneous and isotropic background plus scalar perturbations, with metric:
\begin{align}
\mathrm ds^2=a^2\left[-(1+2\Psi)\mathrm d\eta^2+ (1-2\Phi)\delta_{ij}\mathrm dx^j\mathrm dx^j\right]\,.
\end{align}
Here $a$ is the scale factor, $\eta$ denotes conformal time, and the two gravitational potentials $\Psi$ (time distortion) and $\Phi$ (spatial distortion) denote the perturbations of the geometry. Without assuming any theory of gravity, we then encode the evolution of the Weyl potential, $\Psi_W = (\Phi+\Psi)/2$, from redshift $z_*$ to redshift $z$ into the free function $\hJ(z)$ defined through
\begin{align}
\Psi_W(k,z)= \left(\frac{\HH(z)}{\HH(z_*)} \right)^2 \sqrt{\frac{B(k,z)}{B(k,z_*)}}\hJ(z)\frac{\Psi_W(k,z_*)}{\sigma_8(z_*)}\, ,
\label{eq:hatJ}
\end{align}
where $\HH$ is the Hubble parameter in conformal time and $B$ is a boost factor encoding non-linearities.
We fix $z_*=10$, well before the acceleration of the Universe started, and we assume that at that redshift gravity behaves as in GR. This is motivated by the fact that modifications of gravity are proposed as a solution to explain the accelerated expansion of the Universe, and as such, they are expected to play a role at late times, during the acceleration. In contrast, early-time observations are fully consistent with GR, motivating the hypothesis that any deviation should be strongly suppressed at early times. Note that with this assumption, we automatically exclude early dark energy models, which would lead to modifications at $z_*$. 

In this set-up, all quantities at $z_*$ are given by their GR predictions and are constrained by CMB observations. The function $\hJ$, on the other hand, is left free and will be constrained by galaxy-galaxy lensing measurements. In full generality, $\hJ$ can depend on both redshift and wavenumber $k$. In practice, however, current data are not constraining enough to be sensitive to a $k$-dependence. We therefore drop this dependency in our measurement~\footnote{In GR, $\hJ$ is independent of $k$. In theories beyond GR, $\hJ$ generically acquires a scale-dependence. However, this dependence turns out to be very mild inside the horizon (in the quasi-static approximation) for a large variety of theories beyond GR~\cite{Gleyzes:2015rua,Pogosian:2021mcs}.}. The boost $B(k,z)$ accounts for non-linearities that affect the galaxy-galaxy lensing signal. In this work, we model this boost in GR, accounting for the impact of $w_0$ and $w_a$. Ideally, one would want to have a model-independent form for this boost, valid for broad classes of theories beyond GR, but this is highly non-trivial and no such framework has
been developed yet\,\footnote{Various approaches have been explored in the literature to incorporate non-linear modified gravity effects while encompassing a large class of gravity theories. Recently, Ref.~\cite{Srinivasan:2026fxo} has emulated the non-linear matter power spectrum using the phenomenological $\mu-\Sigma$ parametrization, building up on previous work~\cite{Thomas:2020duj, Srinivasan:2021gib, Srinivasan:2023qsu, Srinivasan:2024nkv}. Another approach is the extension of the effective field theory of dark energy to non-linear scales, see e.g.~\cite{Frusciante:2017nfr, Cusin:2017mzw, Yamauchi:2017ibz}. However, there is no unique mapping between these phenomenologically or effective theory motivated parameters and our observable $\hat{J}$. The application of such non-linear approaches to our work is thus non-trivial and will be further investigated in the future.}.  
What is important, however, is that the boost cannot create a fictitious tension. If our Universe is truly governed by GR, the modelling of the boost and thus the measurements of $\hJ$ will be correct, and $\hJ$ should agree with the GR theoretical prediction. As a consequence, if we detect a tension between the measured $\hJ$ and its prediction in GR, it cannot be caused by the boost and thus truly indicates physics beyond GR. 

As shown in~\cite{Tutusaus:2023aux}, the galaxy-galaxy lensing angular power spectra can be written in terms of $\hJ$ as 
\begin{align}
&C_{\ell}^{\Delta\kappa}(z_i,z_j)=\frac{3}{2}\hat{b}(z_i)\hat{J}(z_i)\int\text{d}z\,n_i(z)\mathcal{H}^2(z)\label{eq:DeltakappaJ}\\
&\times B\left(k_{\ell},\chi\right)\frac{P_{\delta\delta}^{\rm lin}\left(k_{\ell},z_*\right)}{\sigma_8^2(z_*)}
    \int\text{d}z'n_j(z')\frac{\chi'(z')-\chi(z)}{\chi(z)\chi'(z')}\,, \nonumber
\end{align}
where $n_i$ and $n_j$ denote the galaxy distribution function of the lenses and sources, respectively, $\hat{b}(z_i)=b(z_i)\sigma_8(z_i)$ with $b$ the linear galaxy bias, and $k_\ell\equiv (\ell+1/2)/\chi$, with $\chi$ the comoving distance. Note that here we have assumed that $\hat{b}(z)$ and $\hat{J}(z)$ evolve slowly with redshift, such that we can take them out of the integral and evaluate them at the mean redshifts $z_i$ of the lens distributions. 

For the galaxy clustering angular power spectra, we use the Limber approximation for $\ell>200$ and the full-sky expression for smaller $\ell$:
\begin{align}\label{eq:DeltaDeltaJ}
&C_{\ell}^{\Delta\Delta}(z_i,z_j)=\hat{b}(z_i)\hat{b}(z_j)\int\text{d}z\,n_i(z)n_j(z)\frac{\mathcal{H}(z)(1+z)}{\chi^2(z)}\nonumber\\
&\quad\times B\left(k_{\ell},\chi\right)\frac{P_{\delta\delta}^{\rm lin}\left(k_{\ell},z_*\right)}{\sigma_8^2(z_*)}\,, \quad \mbox{for $\ell\geq 200$}\,.\\
&C_{\ell}^{\Delta\Delta}(z_i,z_j)= 
\frac{2}{\pi}\hat{b}(z_i)\hat{b}(z_j)\! \int \! \text{d}\chi_1\, n_i(\chi_1)(1+z(\chi_1))\HH(\chi_1)\nonumber\\
&\quad\times\int \text{d}\chi_2\, n_j(\chi_2)(1+z(\chi_2))\HH(\chi_2)\\
&\quad\times\int_0^{\infty}\text{d}k\, k^2 \frac{P_{\delta\delta}^{\rm lin}(k,z_*)}{\sigma^2_8(z_*)}j_{\ell}(k\chi_1)j_{\ell}(k\chi_2)\,,
     \quad \mbox{for $\ell< 200$}\,.\nonumber
\end{align}
The two signals $C_\ell^{\Delta\kappa}$ and $C_\ell^{\Delta\Delta}$ can therefore be expressed uniquely in terms of: (1) the initial matter power spectrum at $z_*$; (2) the function $\hJ(z_i)$ at the redshift of the lenses; (3) the function $\hat{b}(z_i)$ at the redshift of the lenses; and (4) the Hubble parameter $\HH(z)$ and comoving distance $\chi(z)$.

Our goal is to constrain together the cosmological parameters that impact the initial power spectrum, namely $\Omega_{\rm m}, \Omega_{\rm b}, n_s, A_s$ and $h$, the functions $\hJ$ and $\hat{b}$ that govern the evolution of the galaxy density and Weyl potential, and the parameters $w_0$ and $w_a$ that impact $\HH(z)$ and the distance $\chi(z)$.

\section{Datasets and methods}
\label{sec:data}

We use the DES Y3 data to constrain $\hat{b}$ and $\hat{J}$. Following the baseline DES Y3 $2\times2$pt analysis~\cite{DES:2021bpo}, we use the \texttt{METACALIBRATION} sample of source galaxies and \texttt{MagLim} lens galaxy sample, and we remove the two highest redshift bins in \texttt{MagLim} due to potential residual systematics. We have therefore four free values of $\hat{J}(z_i)$ and $\hat{b}(z_i)$ corresponding to the four redshift bins of the lens galaxy sample, at effective redshifts $z_i=
\{0.295,0.467,0.626,0.771\}$. We model intrinsic alignment (IA) using the non-linear alignment (NLA) model with two free nuisance parameters for the amplitude $A_{\rm IA}$ and redshift scaling $\eta_{\rm IA}$. We adopt the same nuisance templates for the lens and source redshift distributions and shear calibration, as well as the recommended nuisance priors, as DES Y3 baseline~\cite{DES:2021bpo}. We consider the DES Y3 2x2pt data consisting of galaxy clustering and galaxy-galaxy lensing, and we use the same scale cuts as in the DES Y3 baseline analysis. In~\cite{Tutusaus:2023aux}, more stringent scale cuts were as well explored, finding that while the uncertainties on $\hJ$ increase when 
removing scales, the mean values are almost not affected.

The evolution of both background and perturbations contributes non-trivially to the theoretical prediction of $\hat{J}$. Therefore, to obtain well-measured $\hat{J}$ and estimate the potential tension with the theoretical GR prediction, it is necessary to have precise constraints on the background evolution. In \cite{Tutusaus:2023aux}, this was done by applying Gaussian CMB priors on cosmological parameters, informative to both the background evolution and the initial condition of perturbations. This is, however, inconsistent in the presence of evolving dark energy because the Gaussian priors are obtained assuming $\Lambda$CDM. In this paper, we therefore use the full CMB temperature and polarization auto and cross spectra measured by Planck 2018. More precisely, we use the \texttt{plik\_lite} high-$\ell$ TTTEEE data combined with the \texttt{SimAll} low-$\ell$ EE data, with a single nuisance parameter $A_{\rm planck}$ \cite{Planck:2018vyg}. The low-$\ell$ EE data are included because they are essential in breaking degeneracy between the effective reionization optical depth $\tau$ and the primordial amplitude $A_s$. Since there is strong astrophysical indication that reionization happens before $z=6.5$~\cite{Fan:2005es}, we can reasonably assume that the constraint on $\tau$ from low-$\ell$ CMB polarization is not impacted by $\hat{J}$. On the contrary, the low-$\ell$ temperature data directly depend on $\hJ$ through the integrated Sachs-Wolfe effect. To be consistent we would need to adapt their modeling, which is beyond the scope of this analysis. Therefore, to be conservative, we do not include the low-$\ell$ temperature data.

CMB lensing depends on the integral of $\hJ$ (from the last scattering surface to today) and, for consistency, should therefore be expressed in terms of this quantity. This would require implementing a smooth version of $\hJ$ to compute the integral, which is non-trivial and beyond the scope of this paper.  Therefore, we do not include the lensing reconstruction (obtained from the CMB trispectrum) in our data set, nor its correlation with galaxy clustering. We, however, still need to account for the lensing contribution to the temperature and polarization spectra, which distorts the primordial CMB signal. For this, we perform two different analyses. In the first (conservative) one, we include the parameter $A_{\rm lens}$ in front of the lensing contribution, and we marginalize over it. This effectively removes the lensing contribution from the temperature and polarization: all information from CMB lensing is transferred into a tight constraint on $A_{\rm lens}$. 
As we will see, in this case, the constraint on the primordial amplitude $A_s$ is slightly shifted and enlarged, due to the removal of the lensing information, which partially contributes to the constraint on $A_s$.
This in turn enlarges the uncertainty on the GR prediction for $\hJ$. To quantify the impact of this loss of information, we therefore perform a second analysis, where we model the CMB lensing in GR. This is obviously not fully consistent. The inconsistency is somewhat mitigated by the fact that the CMB lensing kernel peaks around redshift 2, well above our $\hJ$ measurements. However, due to the broadness of the kernel, there is still non-negligible information coming from lower redshifts. This second analysis should therefore be taken with caution, but it has the advantage of providing a way of assessing the tension in $w_0w_a$CDM models that we could expect in the case where lensing information is consistently modeled and taken into account. 

In addition to CMB spectra, we also include in our analysis the BAO measurement from DESI DR2 \cite{DESI:2025zgx} and luminosity distance measurements from type Ia supernovae compiled in the Pantheon+ sample \cite{Scolnic:2021amr}. These data are necessary to break the degeneracies between $w_0$, $w_a$ and $\Omega_{\rm m}$.

Following~\cite{Tutusaus:2023aux}, we use \texttt{CosmoSIS} \cite{Zuntz:2014csq}, interfaced with the nested sampler \texttt{PolyChord} \cite{Handley:2015fda} for model sampling. To obtain the theoretical predictions of $\hJ$ in GR, we use the cosmological code \texttt{CAMB} \cite{Lewis:1999bs} which computes the background and linear perturbation cosmology.

\section{Results}
\label{sec:results}

\begin{figure*}
    \centering
    \includegraphics[width=0.9\linewidth]{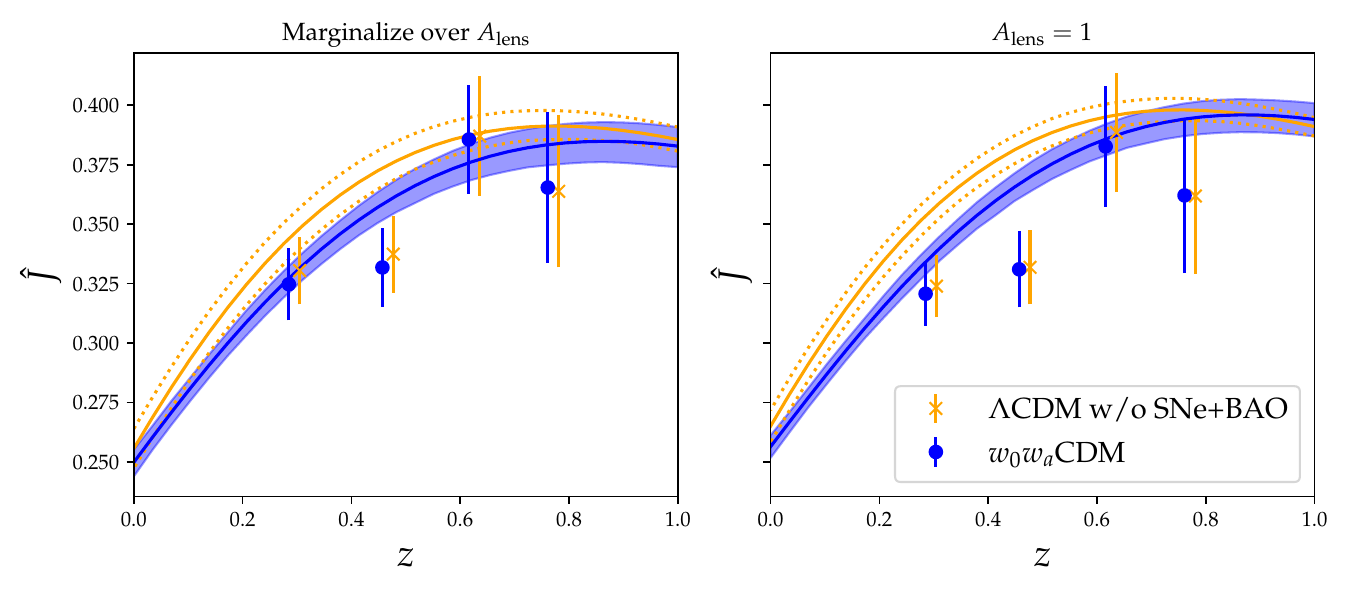}
    \caption{Measured values of $\hJ$ together with 1$\sigma$ uncertainties in $\Lambda$CDM (using DES-Y3+CMB) and in $w_0w_a$CDM (using DES-Y3+CMB+BAO+SNe) at the effective redshifts $z_i=[0.295,0.467,0.626,0.771]$ of the four tomographic bins of DES-Y3. The solid lines show the predictions for $\hJ$ in GR, $\hat{J}_{\rm GR}(z)=\Omega_{\rm m}(z)\sigma_8(z)$, for the corresponding model. The shaded regions indicate the 1$\sigma$ uncertainty on those predictions due to uncertainties in the cosmological parameters. The two panels show the results for the two different analyses: in the left panel, we marginalize over the $A_{\rm lens}$ parameter, while in the right panel, $A_{\rm lens}$ is fixed to 1.  }
    \label{fig:jhatofz}
\end{figure*}

We have tested different combinations of data. For the $\Lambda$CDM case, in our baseline analysis we combine DES Y3 with CMB. We do not include geometrical data from BAO and SNe, since those are in $\sim 2\sigma$ tension with the CMB in $\Lambda$CDM~\cite{DESI:2025zgx}. However, even without BAO and SNe, we find that the background is well constrained in $\Lambda$CDM, since it is mainly controlled by a single parameter, $\Omega_{\rm m}$, when ignoring the subdominant radiation and neutrinos. In contrast, in $w_0w_a$CDM, due to the additional time dependence of dark energy, SNe+BAO data are necessary to constrain the background and reach similar constraining power on $\hat{J}$ as in $\Lambda$CDM. Therefore, the results shown in this paper for $w_0w_a$CDM are obtained using the most constraining datasets DES-Y3+CMB+BAO+SNe unless explicitly mentioned otherwise.

\begin{table*}
\caption{Mean and $1\sigma$ posterior results of the measured $\hat{J}(z_i)$ in $\Lambda$CDM and in $w_0w_a$CDM. The measured $\hat{J}$ from Ref.~\cite{Tutusaus:2023aux} in $\Lambda$CDM with the same DES Y3 data and a CMB prior on cosmological parameters is also included for reference. The last column reports the inconsistency between the measured $\hat{J}(z=0.467)$ (in the second bin) and the corresponding GR prediction, which is the most significant among the four bins, in each model. } 
    \label{tab:jhat}
\renewcommand{\arraystretch}{1.5}
    \begin{tabular}{c|c|cccc|c}
         \multicolumn{2}{c|}{Model}&$\hat{J}(z=0.295)$&$\hat{J}(z=0.467)$&$\hat{J}(z=0.626)$&$\hat{J}(z=0.771)$&Tension  \\
         \hline \hline
         \multirow{2}{*}{Varying $A_{\rm lens}$}
         &$\Lambda$CDM
         &$0.330\pm 0.014            $
         &$0.337\pm 0.016            $
         &$0.387\pm 0.025            $
         &$0.364^{+0.029}_{-0.033}   $
         &$2.1\sigma$
         \\
         &$w_0w_a$CDM
         &$0.325^{+0.014}_{-0.016}   $
         &$0.332\pm 0.016            $
         &$0.386\pm 0.023            $
         &$0.365^{+0.029}_{-0.034}   $
         &$1.6\sigma$    
         \\
         \hline
         \multirow{2}{*}{$A_{\rm lens}=1$}
         &$\Lambda$CDM
         &$0.324\pm 0.013            $
         &$0.332\pm 0.016            $
         &$0.389\pm 0.025            $
         &$0.362\pm 0.033            $
         &$3.1\sigma$ 
         \\
         &$w_0w_a$CDM
         &$0.321\pm 0.013            $
         &$0.331\pm 0.016            $
         &$0.383\pm 0.025            $
         &$0.362\pm 0.033            $
         &$2.2\sigma$ 
         \\
         \hline
         \multicolumn{2}{c|}{Ref.~\cite{Tutusaus:2023aux}}&$0.325\pm0.015$&$0.333^{+0.017}_{-0.019}$&$0.387^{+0.026}_{-0.029}$&$0.354\pm0.035$&$2.8\sigma$
    \end{tabular}
\end{table*}

\begin{figure}
    \centering
    \includegraphics[width=\linewidth]{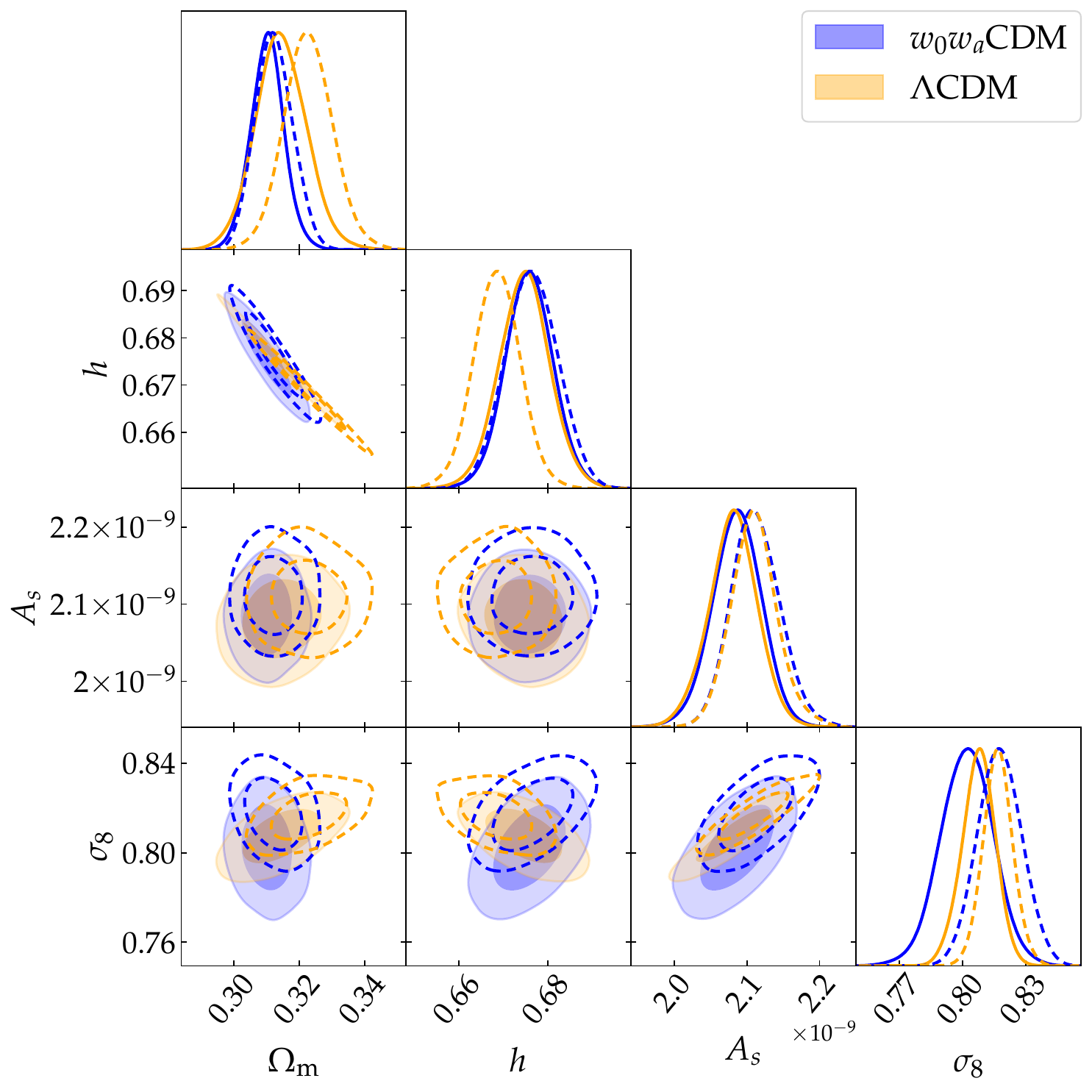}
    \caption{68\% and 95\% posterior distributions of cosmological parameters in the $\Lambda$CDM and $w_0w_a$CDM models. The conservative result ($A_{\rm lens}$ marginalized) is plotted with filled contours and solid lines, while the $A_{\rm lens}=1$ result is shown with dashed lines of the same color.}
    \label{fig:omh0}
\end{figure}

In Fig.~\ref{fig:jhatofz} we show the measured $\hat{J}$ in $\Lambda$CDM and in $w_0w_a$CDM, together with the respective theoretical predictions in GR, for the two different analyses (with and without $A_{\rm lens}$). 
We see that, despite the influence of evolving dark energy on the background, the measured $\hat{J}$ values are very similar in $\Lambda$CDM and in $w_0w_a$CDM, with a shift of only a few percent in the central values between different models and datasets. Our measurements are also highly consistent with the previous result in~\cite{Tutusaus:2023aux}, using CMB priors instead of the full CMB spectra. This very weak dependence of the measured $\hJ$ on the background evolution can be directly understood from the way it enters the $C_\ell$'s in Eq.~\eqref{eq:DeltakappaJ}. While the background evolution does impact $\HH(z)$ and the ratio of comoving distances in the lensing kernel, they only vary by a few percent between $\Lambda$CDM and our posterior constraint on evolving dark energy, namely $-0.855^{+0.054}_{-0.047}$ and $w_a=-0.48\pm0.22$. Consequently, only a tiny shift in $\hat{J}$ is required to compensate this change and keep the measured $C_\ell$'s unchanged\,\footnote{This is in line with the findings of~\cite{Heydenreich:2025xim}, where differences of 7 percent or lower have been found between $\Lambda$CDM and $w_0w_a$CDM, although at the level of galaxy-galaxy lensing angular power spectra rather than inferred parameters.}.  This shows that $\hJ$ is a powerful quantity to capture the perturbed evolution of the geometry, distinctly from its background evolution.

In contrast with the measurements, the theoretical predictions for $\hJ$ in GR are more strongly affected by the background evolution, as we see from Fig.~\ref{fig:jhatofz}. 
Allowing for evolving dark energy, the theoretical prediction decreases by more than 1$\sigma$ at intermediate redshift with respect to the $\Lambda$CDM case, in both analyses. This decrease can be traced to the fact that BAO+SNe+CMB favor an equation of state larger than $-1$ below $z\simeq 0.4$, which decreases the depth of the gravitational potentials in GR due to a stronger acceleration. More precisely, evolving dark energy modifies the evolution of $\Omega_{\rm m}(z)$ which directly impacts $\hJ_{\rm GR}$ (together with $\sigma_8(z)$). At $z=0$ the value for $\hat{J}_{\rm GR}$ in $w_0w_a$CDM  is slightly smaller than in $\Lambda$CDM due to the smaller $\Omega_{\rm m}(z=0)$, as we see from Fig.~\ref{fig:omh0}. Having $w>-1$ at $z\lesssim0.4$ amplifies this difference since it implies that the dark energy density becomes higher in the past, causing the $\Omega_{\rm m}(z)$ in $w_0w_a$CDM to increase more slowly -- going backward in time -- than in $\Lambda$CDM. As a result, $\hat{J}_{\rm GR}$ in $w_0w_a$CDM is consistently smaller than that in $\Lambda$CDM in the intermediate redshift range, reducing the tension. 
For example, in the second redshift bin, which has the largest tension, it goes from 2.2$\sigma$ in $\Lambda$CDM to 1.6$ \sigma$ in $w_0w_a$CDM in the first analysis (marginalize over $A_{\rm lens}$), and from 3.1$\sigma$ in $\Lambda$CDM to 2.1$\sigma$ in $w_0w_a$CDM in the second analysis ($A_{\rm lens}=1$), see Table~\ref{tab:jhat}. At higher redshifts, the effect reverses. Since $w$ becomes smaller than $-1$, the energy density of dark energy decreases and $\Omega_{\rm m}(z)$ increases with respect to $\Lambda$CDM. As a consequence, $\hJ_{\rm GR}$ increases more steeply than in $\Lambda$CDM in the past, and the two curves agree around $z=1$. In addition to this effect, $\hJ_{\rm GR}$ is also, in principle, affected by changes in $\sigma_8(z)$ due to evolving dark energy. We find, however, that these changes are subdominant with respect to the ones due to the evolution of $\Omega_{\rm m}(z)$. 

From the two analyses, we see that evolving dark energy with a phantom crossing reduces the tension in the Weyl potential by decreasing the depth of the potentials with respect to $\Lambda$CDM at intermediate redshift. The significance of the tension, both in $\Lambda$CDM and in $w_0w_a$CDM, varies, however, when we include or not $A_{\rm lens}$. When $A_{\rm lens}$ is included, the theoretical prediction in GR is about 1$\sigma$ lower than when $A_{\rm lens}$ is fixed. This can be traced back to the mean of the primordial amplitude parameter $A_s$ and matter fraction $\Omega_{\rm m}$, which are shifted to lower values when lensing is removed through $A_{\rm lens}$, as can be seen from Fig.~\ref{fig:omh0}. Moreover, the uncertainty on $\hJ_{\rm GR}$ increases by $\sim$50 percent when $A_{\rm lens}$ is included, due to the increased uncertainty on $A_s$ and $\Omega_{\rm m}$. As a consequence, both in $\Lambda$CDM and in $w_0w_a$CDM, the tension is reduced when $A_{\rm lens}$ is included. For example, in the second redshift bin, the tension in $\Lambda$CDM decreases from 3.1$\sigma$ to 2.1$\sigma$ when including $A_{\rm lens}$ and in $w_0w_a$CDM from 2.2$\sigma$ to 1.6$\sigma$. This range of values is directly linked to our imperfect modelling of CMB lensing: in the first analysis, we artificially remove all information from lensing, by including the ad-hoc parameter $A_{\rm lens}$, while in the second analysis we model CMB lensing in GR. 
In a forthcoming paper, we will model CMB lensing in terms of $\hJ$ and include also the lensing reconstruction in our analysis, in order to assess how CMB lensing information impacts $\hJ$ compared to shear measurements from DES.

\section{Discussion and interpretation}
\label{sec:discussion}

\begin{figure*}[t]
    \centering
    \includegraphics[width=0.8\textwidth]{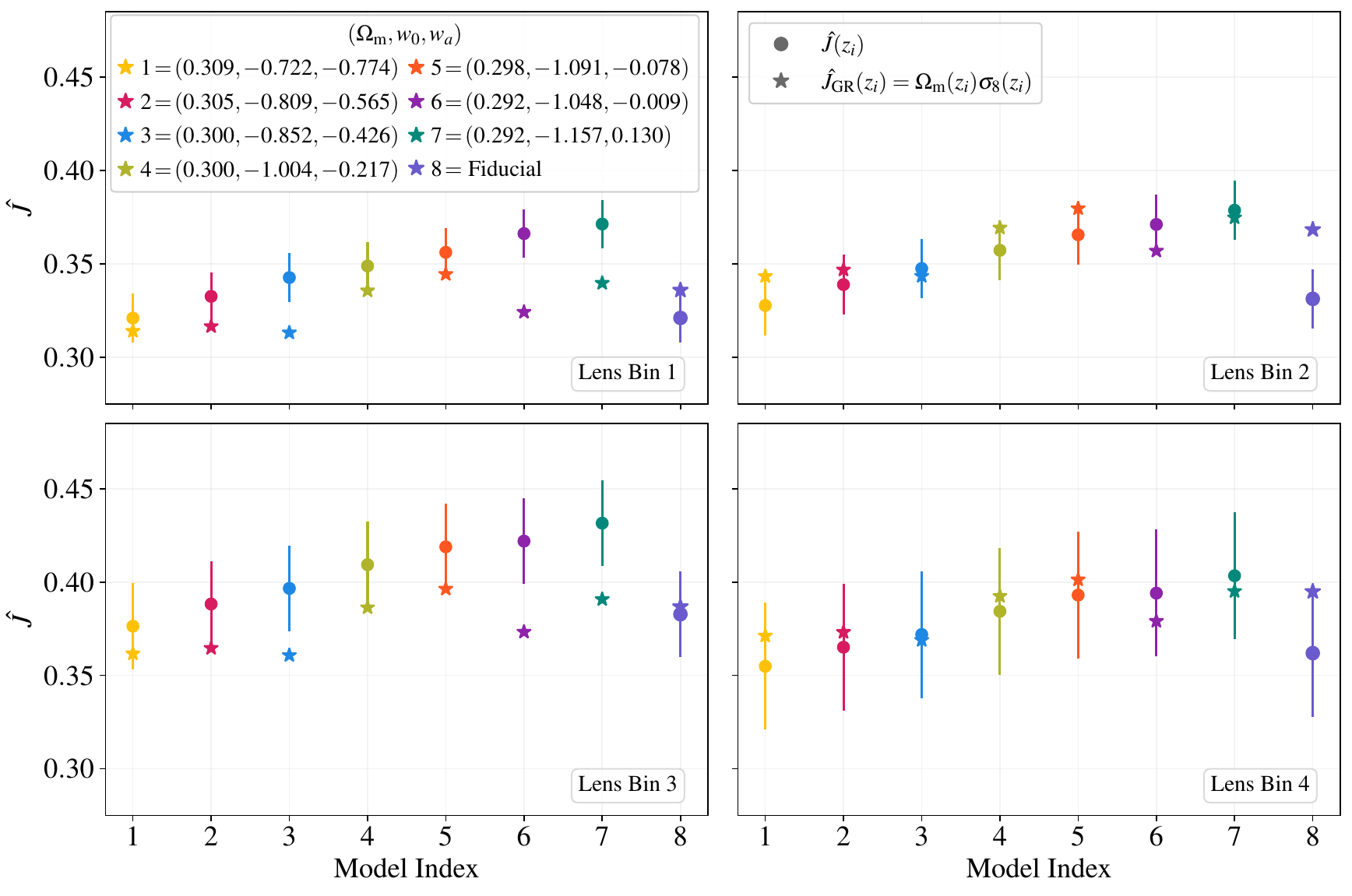}
    \caption{We show the values of the GR prediction $\hat{J}_{\rm GR}(z_i)$ (stars) and the measured $\hat{J}(z_i)$ (dots) for a representative subset of the selected $w_0w_a$CDM models (with $\chi^2 \leq 1$ in the second bin, as shown in Fig.~\ref{fig:chi2}). The uncertainties are directly drawn from our MCMC analysis, since they are not expected to depend on the choice of cosmological parameters. }
    \label{fig:jhat4bins}
\end{figure*}

As we have seen from Fig.~\ref{fig:jhatofz}, $w_0$ and $w_a$ impact the amplitude and the shape of the GR prediction for $\hJ$. In our two analyses, the tension is reduced due to the decrease of $\hJ_{\rm GR}$ at intermediate redshifts induced by $w_0$ and $w_a$. It does, however, not fully disappear in $w_0w_a$CDM, with a value ranging from 1.6$\sigma$ to 2.2$\sigma$ in the second bin, depending on our treatment of CMB lensing.

Given these results, we aim to assess whether the persistence of the tension reflects an intrinsic limitation of $w_0w_a$CDM  models -- namely, its inability to simultaneously reproduce the background evolution measured by DESI, and the evolution of the perturbed geometry encoded in the  Weyl potential -- or whether it is instead driven by the particular best-fit model selected by the data, which yields a value of $\hJ_{\rm GR}$ that is higher than observed. To answer this question, we concentrate on a reduced set of data, consisting of the galaxy-galaxy lensing spectra, $C^{\Delta\kappa}_\ell$ (which constrain the Weyl potential), and on the angular diameter distances, $D_{\rm M}$, and the Hubble distances, $D_{\rm H}$, measured by DESI. We search for models that are consistent with this subset of data while simultaneously removing the tension in the Weyl potential. From our results in the previous section, we already know that such models cannot be a good fit to the full set of data since otherwise they would have been identified in our MCMC analysis. However, the goal of this section is precisely to explore models outside of the MCMC constraints by removing some of the data (CMB, SNe and galaxy clustering) in order to allow for more drastic changes in $\hJ$.

When moving away from the fiducial (i.e.\ the mean of our MCMC analysis), we keep the parameters $\Omega_{\rm b}$, $A_s$, $n_s$, $h$ and $\tau$ fixed, and we only vary $w_0$, $w_a$ and $\Omega_{\rm m}$, since they have the strongest impact on the theoretical prediction for $\hJ$. While $A_s$ also impacts $\hJ$ in a non-negligible way, varying it does not modify the shape of $\hJ$ but only its overall amplitude. Here, we focus on exploring models with varying redshift behaviors, fitting $\hJ$ well across all redshifts, and we thus keep $A_s$ fixed. To remain close to DESI's fiducial, we impose the constraint that the set of points $\{\Omega_{\rm m}, w_0, w_a\}^{\rm DESI}$ is consistent with one of DESI's distance measurements. Specifically, we choose to reproduce the ratio $(D_{\rm M}/D_{\rm H})_{\rm fiducial}$ at DESI's effective redshift $z_{\rm eff}=0.934$ (where the fiducial prediction is in excellent agreement with the measured ratio). By taking the ratio instead of the individual distances, our procedure does not depend on $h$. We find, therefore, a surface $\{\Omega_{\rm m}, w_0, w_a\}^{\rm DESI}$ around the fiducial model, which agrees with one data point but can in principle deviate at other redshifts, allowing us to explore diverse expansion histories.

\begin{figure*}[t]
    \centering
     \includegraphics[width=0.49\textwidth]{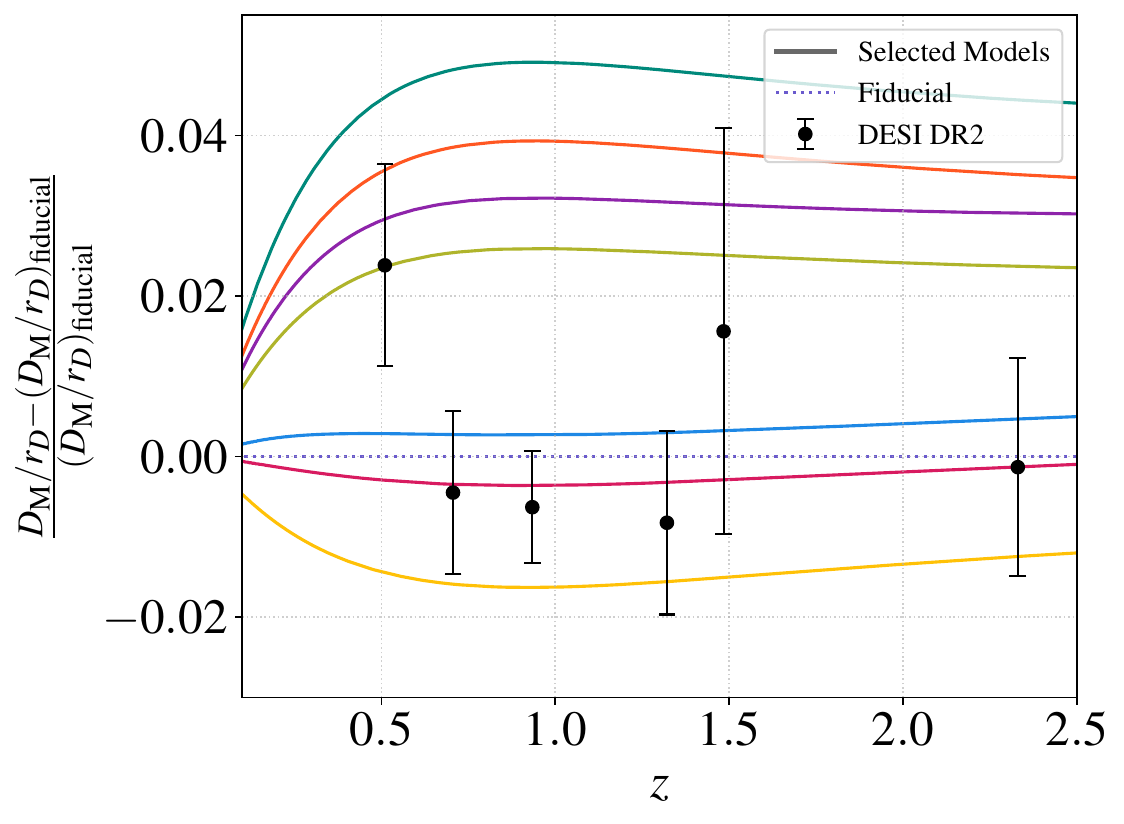}
     \includegraphics[width=0.49\textwidth]{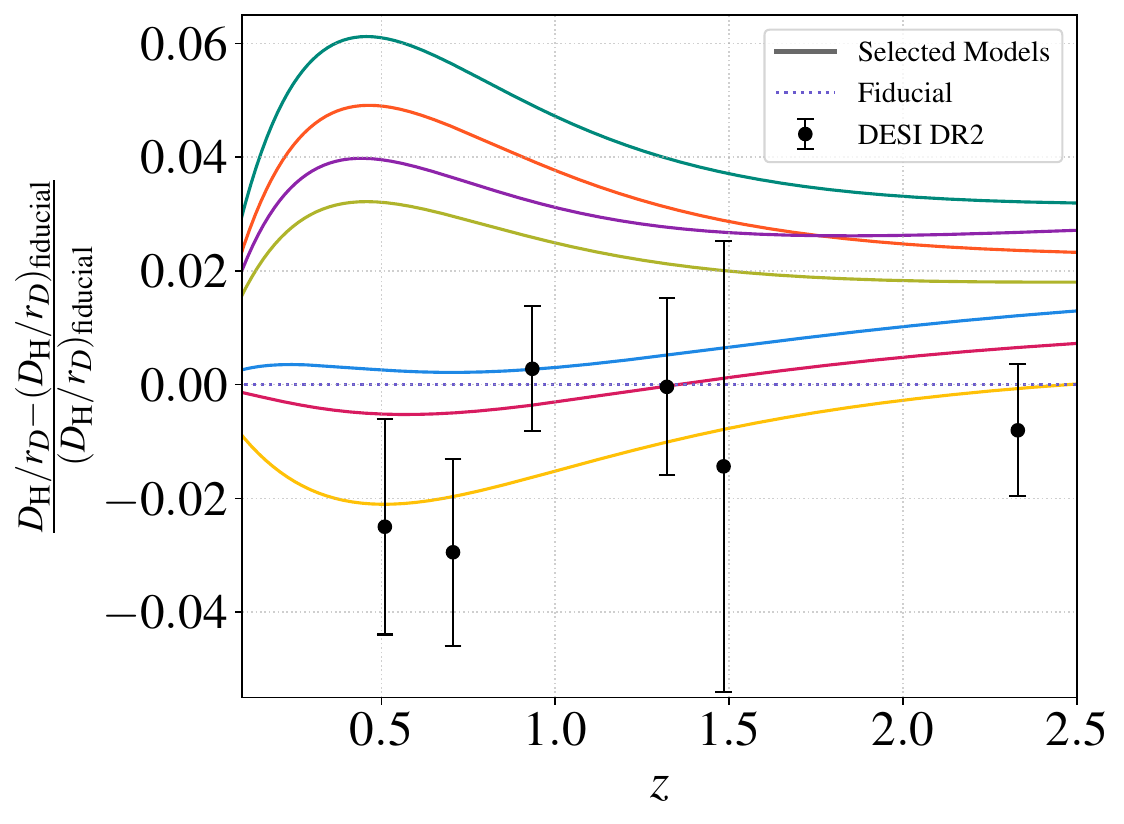}
    \caption{Angular diameter distance $D_{\rm M}$ (left panel) and radial distance $D_{\rm H}$ (right panel) measured by DESI DR2, together with their 1$\sigma$ uncertainty. We normalize the distances by the sound horizon $r_D$, and we show the difference with respect to the fiducial value. The colored lines show the theoretical prediction for a selection of models that solve the $\hJ$ tension in the second bin (same color coding as in Fig.~\ref{fig:jhat4bins}). Note that when exploring models away from the fiducial, we keep $r_D$ fixed since it is tightly constrained by the CMB.}
    \label{fig:dHdM}
\end{figure*}

Our goal is then to determine if there is a subset of models on the selected surface $\{\Omega_{\rm m}, w_0, w_a\}^{\rm DESI}$ that would fully resolve the $\hJ$ tension. To achieve this goal, we discretize the surface $\{\Omega_{\rm m}, w_0, w_a\}^{\rm DESI}$ and for each point we fit for $\hJ$ using Eq.~\eqref{eq:DeltakappaJ} with a proxy for the measured $C_\ell^{\Delta\kappa}$\,\footnote{Ideally, we should use the measured $C_\ell^{\Delta\kappa}$ in our fit. However, we do not have access to it, first, since DES's public release provides the correlation function rather than the angular power spectra, and second, since only the full signal is measured, which contains redshift-space distortions, magnification, and intrinsic alignment. Therefore, in this analysis, as a proxy for the measured $C_\ell^{\Delta\kappa}$, we use the theoretical $C_\ell^{\Delta\kappa}$ at the mean parameters of the MCMC analysis. Note that we could also use the best-fit instead of the mean as a proxy, but we would then infer the best-fit $\hJ$ instead of the mean $\hJ$, which can differ from Fig.~\ref{fig:jhatofz}, thus complicating the comparison.  For the uncertainty, we use the theoretical Gaussian covariance.}. This inferred $\hJ$ provides a good approximation of the measured $\hJ$ that we would infer directly from the data through an MCMC, but under the constraints that the other parameters are fixed, which is sufficient for our discussion. We then compute the GR prediction at the effective redshifts of DES,
$\hat{J}_{\rm GR}(z_i) =\Omega_{\rm m}(z_i)\sigma_8(z_i)$, for each point on the surface $\{\Omega_{\rm m}, w_0, w_a\}^{\rm DESI}$. Next, we select the subset of models that solve the tension in the second redshift bin, i.e.\ such that $\hat{J}_{\rm GR}$ is consistent, within 1$\sigma$, with the ``measured'' $\hat{J}$ inferred from $C_\ell^{\Delta\kappa}$. Note that for this selection, we use the 1$\sigma$ uncertainties from our MCMC analysis (which are not expected to depend much on the cosmology), rather than the ones from our fit. The latter are indeed significantly smaller since all other parameters are kept fixed in the fit.

We find a set of models that fully solve the tension in the second bin, i.e., such that $\chi^2\leq 1$ (see Fig.~\ref{fig:chi2} in Appendix~\ref{app:AddPlots}). These models either solve the tension in the second bin by lowering the GR prediction, or by increasing the measured value, or both, as can be seen from the second panel of Fig.~\ref{fig:jhat4bins}. For example, Model number~1 (yellow) has very little impact on the measured $\hJ$, but it lowers the GR prediction in the second bin from 0.360 to 0.343. Looking at the equation of state for this model (Fig.~\ref{fig:w_z} in Appendix~\ref{app:AddPlots}), we see that it crosses the phantom divide at a larger redshift and remains above the fiducial. As a consequence, $\Omega_{\rm m}(z)$ increases more slowly toward higher redshift (see Fig.~\ref{fig:Om_z} in Appendix~\ref{app:AddPlots}), which lowers $\hJ_{\rm GR}$ and solves the tension in the second bin. From Fig.~\ref{fig:jhat4bins}, we see that, in this Model 1, the GR prediction agrees with the measured $\hJ$ in the other redshift bins as well. However, this model is in tension with the distances measured by DESI, in particular the measured angular diameter distance $D_{\rm M}$, is systematically larger than the prediction in Model 1, as can be seen from the right panel of Fig.~\ref{fig:dHdM}.

At the other extreme, we see that Model number 7 (dark green) solves the tension in the second bin by increasing the measured $\hJ$ from 0.331 to 0.379. This is achieved through an equation of state which is below $-1$ until today (see Fig.~\ref{fig:w_z} in Appendix~\ref{app:AddPlots}). Thus, while the measurement of $\hJ$ has shown very little dependence on the background model in Fig.~\ref{fig:jhatofz}, certain choices of parameters, such as in Model~7 that is far away from the fiducial, can cause a stronger dependence. In particular, keeping the ratio $D_{\rm M}(z)/D_{\rm H}(z)$ fixed at $z_{\rm eff}=0.934$ for this model requires a small $\Omega_{\rm m}=0.292$, well below the fiducial. This set of parameters changes the evolution of $H(z)$ and of the distances, which in turn increases the measured $\hJ$, solving the tension in the second bin. For this model, we see, however, that the tension is shifted to the first and third bins (see Fig.~\ref{fig:jhat4bins}), where the measured $\hJ$ is larger than for the fiducial and now in tension with the GR prediction, which is below. From Fig.~\ref{fig:dHdM} we see moreover that the distances in this Model 7 are always above the measured ones.

In between these two extreme cases, we find models that solve the tension in all bins and are also in relatively good agreement with the measured distances, for example, Model 2 (red). This model lowers the GR predictions while slightly increasing the measured values, and is thus in agreement in all bins. From Fig.~\ref{fig:dHdM}, we see that it is also largely in agreement with both distances.

Our exploration shows, therefore, that, when removing constraints from the CMB, galaxy clustering, and SNe, $w_0w_a$CDM models provide enough freedom to reproduce both the evolution of the Weyl potential in GR (see also Fig.~\ref{fig:JGR_z} in Appendix~\ref{app:AddPlots}) and the evolution of the distances measured by DESI. This behavior is not due to an increase in the uncertainties (that we, on purpose, kept fixed in our exploration), but truly to a different evolution of the Weyl potential and of $H(z)$, which reconciles the measurements with the GR predictions. However, our MCMC analysis shows that -- when combining all data sets -- such models are not favored and that a mild tension in the Weyl potential remains.

\section{Conclusion}
\label{sec:conclusion}

The goal of our study was to understand whether evolving dark energy can solve the tension between the Weyl potential, measured from DES data, and its GR prediction. For this, we first repeated the analysis of~\cite{Tutusaus:2023aux} in $\Lambda$CDM, but combining DES~Y3 data with the full CMB likelihood rather than using only CMB priors. We recovered a tension between $2.1\sigma$ and $3.1\sigma$ in the second DES redshift bin, depending on the treatment of CMB lensing. We then considered $w_0w_a$CDM models, including as well BAO+SNe data, in order to obtain a similar constraining power. Our analysis showed that in this case, the tension is reduced with respect to the $\Lambda$CDM case, but does not fully disappear. More precisely, the residual tension in the second DES redshift bin is at the level of $1.6-2.2\sigma$.

We emphasize that the reduction of the tension in $w_0w_a$CDM arises due to a change in the theory prediction of $\hJ_{\rm GR}$. The measured $\hJ$ values, on the other hand, are only very mildly affected by a change of the background model. This is because $\hJ$ is primarily constrained through cosmological data at the level of perturbations, and the background evolution enters only subdominantly through the lensing kernels.  The lack of a strong dependence between parameters capturing deviations from GR at the level of background versus perturbation is, in fact, not surprising: a similar conclusion can be drawn from the modified gravity constraints of DESI clustering data reported in~\cite{Ishak:2024jhs}, where the phenomenological $\mu-\Sigma$ parameterization was used. The constraints obtained on these parameters, capturing deviations from GR at the level of perturbations, as well vary little between the $\Lambda$CDM and $w_0w_a$CDM cases.

Exploring models away from the MCMC output, by removing the constraining power from CMB, galaxy clustering, and SNe, we found that $w_0w_a$CDM models
have sufficient freedom to reproduce the observed evolution of the Weyl potential and the distance measurements simultaneously. However, these solutions are not favored by the full data set, which instead prefers a lower value of $\hJ$ than that predicted by GR in the second redshift bin. More precise measurements of the Weyl potential are clearly needed to reach decisive conclusions. But if the tension persists, our analysis suggests that some further fundamental modifications to our cosmological model are required to lower the value of $\hJ$, by impacting the Weyl potential not only through the background evolution, but also directly through modifications of gravity or new forces in the dark sector. This could, at the same time, naturally explain the phantom crossing seen by DESI, since it is difficult to obtain such a behavior with dark energy~\cite{Vikman:2004dc,Carroll:2003st}, while modifications of gravity can naturally produce such a crossing~\cite{Ye:2024ywg,Chudaykin:2024gol,Akarsu:2024eoo,Pan:2025psn,Pan:2025psn,Wolf:2025jed,Ye:2026yqk}.

One way to progress further is to add the CMB lensing reconstruction to our analysis, as well as its cross-correlation with galaxy clustering. This would provide additional constraints on the Weyl potential. This will require adapting the formalism by binning $\hJ$ over redshifts and constructing a smooth function which allows integration over $\hJ$. Such a function, while requiring additional modelling choices, could also improve this analysis using DES data, first because it would allow us to model the lensing distortion of the primordial CMB spectrum properly, removing the need for the parameter $A_{\rm lens}$; and second because it would allow us to include the shear auto-correlation. 

In addition, measuring the Weyl potential with upcoming data from surveys such as Euclid\,\footnote{\url{https://www.euclid-ec.org}} or the Legacy Survey of Space and Time\,\footnote{\url{https://rubinobservatory.org}} from the Vera C. Rubin Observatory will be extremely useful thanks to the larger number of photometric bins where $\hJ$ can be measured. The four measurements from DES data 
already show that evolving dark energy does not fully remove the tension, but having more data over a larger range of redshift will definitely help determine the origin of the tension, should it persist in these future data sets.

\section*{Acknowledgments}

B.R., G.Y., M.B. and C.B. acknowledge support from the Swiss National Science Foundation. M.B. acknowledges support from the Tomalla foundation. N.G. acknowledges the support of the Royal Society as a Newton International Fellow (NIF\textbackslash R1\textbackslash 252792) and by the STFC (ST/B001175/1). IT acknowledges support from the Spanish Ministerio de Ciencia, Innovaci\'on y Universidades, projects PID2022-141079NB, PID2022-138896NB; the European Research Executive Agency HORIZON-MSCA-2021-SE-01 Research and Innovation programme under the Marie Sk\l odowska-Curie grant agreement number 101086388 (LACEGAL) and the programme Unidad de Excelencia Mar\'{\i}a de Maeztu, project CEX2020-001058-M. IT has been supported by the Ramon y Cajal fellowship (RYC2023-045531-I) funded by the State Research Agency of the Spanish Ministerio de Ciencia, Innovaci\'on y Universidades, MICIU/AEI/10.13039/501100011033/, and Social European Funds plus (FSE+). 

\appendix
\section{Priors and posterior constraints on the parameters}

\begin{table}[h]
\caption{Prior choice for the sample parameters. $\mathcal{U}$ for uniform prior and $\mathcal{N}$ for Gaussian prior.}
    \label{tab:prior}
    \centering
    \renewcommand{\arraystretch}{1.5}
    \begin{tabular}{c|c}
         Parameter Name& Prior  \\
         \hline \hline
         $\Omega_{\rm m}$& $\mathcal{U}(0.1, 0.9)$\\
         $h$&$\mathcal{U}(0.55,0.91)$\\
         $\Omega_{\rm b}$&$\mathcal{U}(0.03,0.07)$\\
         $10^{9}A_s$&$\mathcal{U}(0.5,5)$\\
         $n_s$&$\mathcal{U}(0.87,1.07)$\\
         $\tau$&$\mathcal{U}(0.01, 0.8)$\\
         $w_0$&$\mathcal{U}(-2,-1/3)$\\
         $w_a$&$\mathcal{U}(-3,0.5)$\\
         $A_{\rm lens}$&$\mathcal{U}(0.5,2)$\\
         \hline
         $\hat{J}_{1,2}$&$\mathcal{U}(0.1,0.6)$ \\
         $\hat{J}_{3,4}$&$\mathcal{U}(0.15,0.65)$ \\
         $\hat{b}_{1,2,3,4}$&$\mathcal{U}(0.1,2.0)$ \\
         \hline
         $A_{\rm planck}$&$\mathcal{N}(1,0.0025)$\\
         \hline
         $M_B$&$\mathcal{U}(-21,-18)$\\
    \end{tabular}
\end{table}

\begin{table*}
\caption{Mean and $1\sigma$ posterior constraints on the cosmological parameters.}
    \label{tab:cosmo_par}
    \centering
    \renewcommand{\arraystretch}{1.5}
    \begin{tabular}{c|cc|cc}
        & \multicolumn{2}{c}{Varying $A_{\rm lens}$}& \multicolumn{2}{|c}{$A_{\rm lens}=1$}\\
        \hline \hline
         &$\Lambda$CDM& $w_0w_a$CDM & $\Lambda$CDM & $w_0w_a$CDM  \\
         \hline
         $\Omega_m$&$0.3144\pm 0.0079          $
                   &$0.3103\pm 0.0052          $
                   &$0.3228\pm 0.0075          $
                   &$0.3123\pm 0.0057          $
                   \\
         $h$&$0.6747\pm 0.0057          $
            &$0.6758\pm 0.0054          $
            &$0.6685\pm 0.0053          $
            &$0.6763\pm 0.0059          $
            \\
         $\Omega_b$&$0.04936\pm 0.00064        $
                   &$0.04946\pm 0.00084        $
                   &$0.04993\pm 0.00062        $
                   &$0.04906\pm 0.00086        $
                   \\
         $10^9A_s$&$2.081\pm 0.034$
                  &$2.086\pm 0.034$
                  &$2.110\pm 0.032$
                  &$2.112^{+0.031}_{-0.035}$
                  \\
         $n_s$&$0.9664\pm 0.0043          $
              &$0.9713\pm 0.0040          $
              &$0.9627\pm 0.0040          $
              &$0.9665\pm 0.0033          $
              \\
         $\tau$&$0.0498\pm 0.0077          $
               &$0.0525\pm 0.0077          $
               &$0.0548^{+0.0065}_{-0.0078}$
               &$0.0574^{+0.0072}_{-0.0081}$
               \\
         \hline
         $w_0$&-
              &$-0.855^{+0.054}_{-0.047}  $
              &-
              &$-0.830^{+0.051}_{-0.060}  $
              \\
         $w_a$&-
              &$-0.48\pm 0.22             $
              &-
              &$-0.67^{+0.22}_{-0.19}     $
              \\
         \hline
         $A_{\rm lens}$&$1.139\pm 0.057            $
                       &$1.158^{+0.058}_{-0.065}   $
                       &-
                       &-\\
    \end{tabular}
\end{table*}

In the analysis performed in the main text, for cosmological parameters we sample the matter fraction $\Omega_{\rm m}$, baryon fraction $\Omega_{\rm b}$, Hubble constant $h$, primordial curvature perturbation spectrum amplitude $A_s$ and tilt $n_s$, and the effective optical depth $\tau$ of reionization. For $w_0w_a$CDM we further sample the CPL parameters $w_0$ and $w_a$. Table~\ref{tab:prior} lists the the priors of the sampled parameters. 

Our nuisance parameters include the Planck calibration $A_{\rm planck}$, the absolute magnitude calibration $M_B$ of the SNe data, the amplitude $A_{\rm IA}$ and redshift scaling $\eta_{\rm IA}$ of the NLA model, shear calibration bias ${m_i}$ and the redshift systematics of the photometric samples as the mean shifts $\{\Delta z_l, \Delta z_s\}$ and rescaling $\sigma_{z}$ of the bin redshift distribution. For the redshift systematics of the photometric samples, we adopt the same prior as DES Y3 baseline \cite{DES:2021wwk}. 

The posterior constraints on the cosmological parameters are reported in Table~\ref{tab:cosmo_par}.

\section{Additional plots} \label{app:AddPlots}

\begin{figure}
    \centering
    \includegraphics[width=1\linewidth]{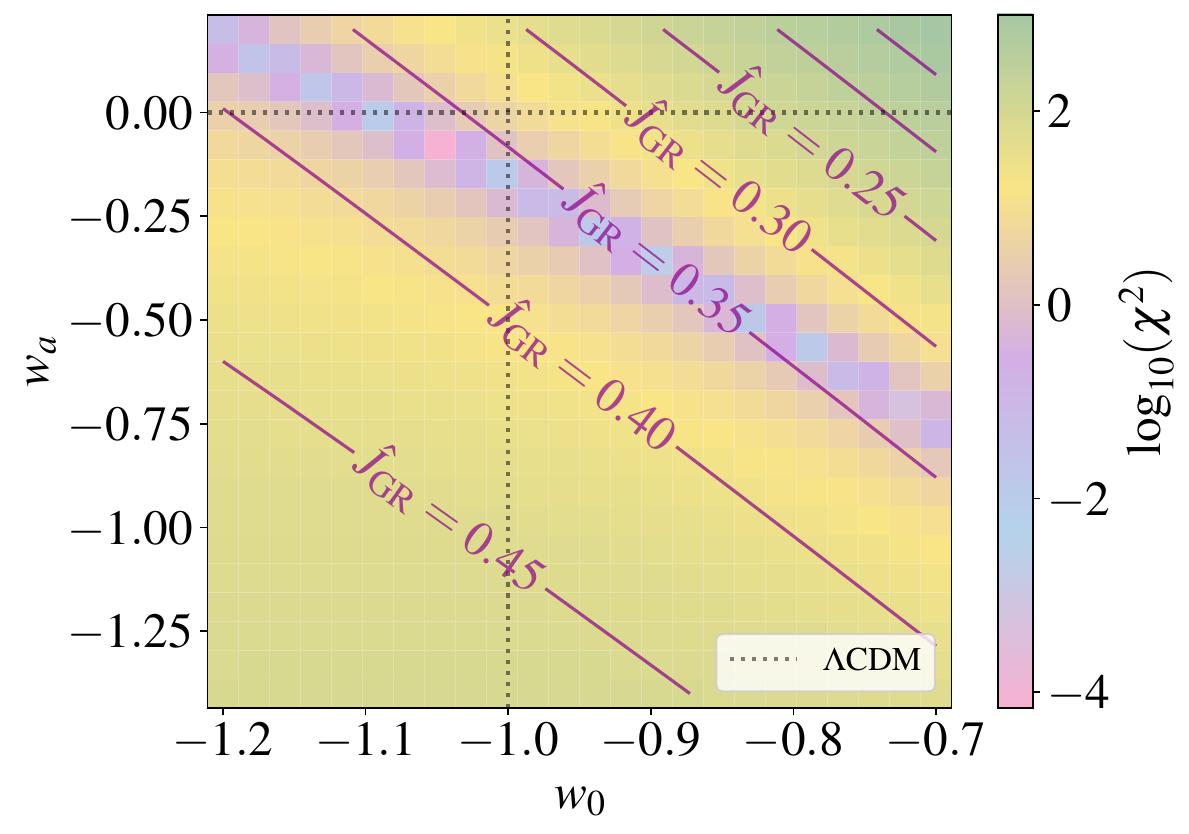}
    \caption{We show $\chi^2=(\hat{J}_{\rm GR} - \hat{J})^2/\sigma^2$ in the second redshift on the surface $\{\Omega_{\rm m}, w_0, w_a\}^{\rm DESI}$, as a function of $w_0$ and $w_a$. The models that solve the tension in the second bin are distributed in the blue-pink region. We also indicate the value of the GR prediction on the surface.}
    \label{fig:chi2}
\end{figure}

\begin{figure}
    \centering
    \includegraphics[width=1\linewidth]{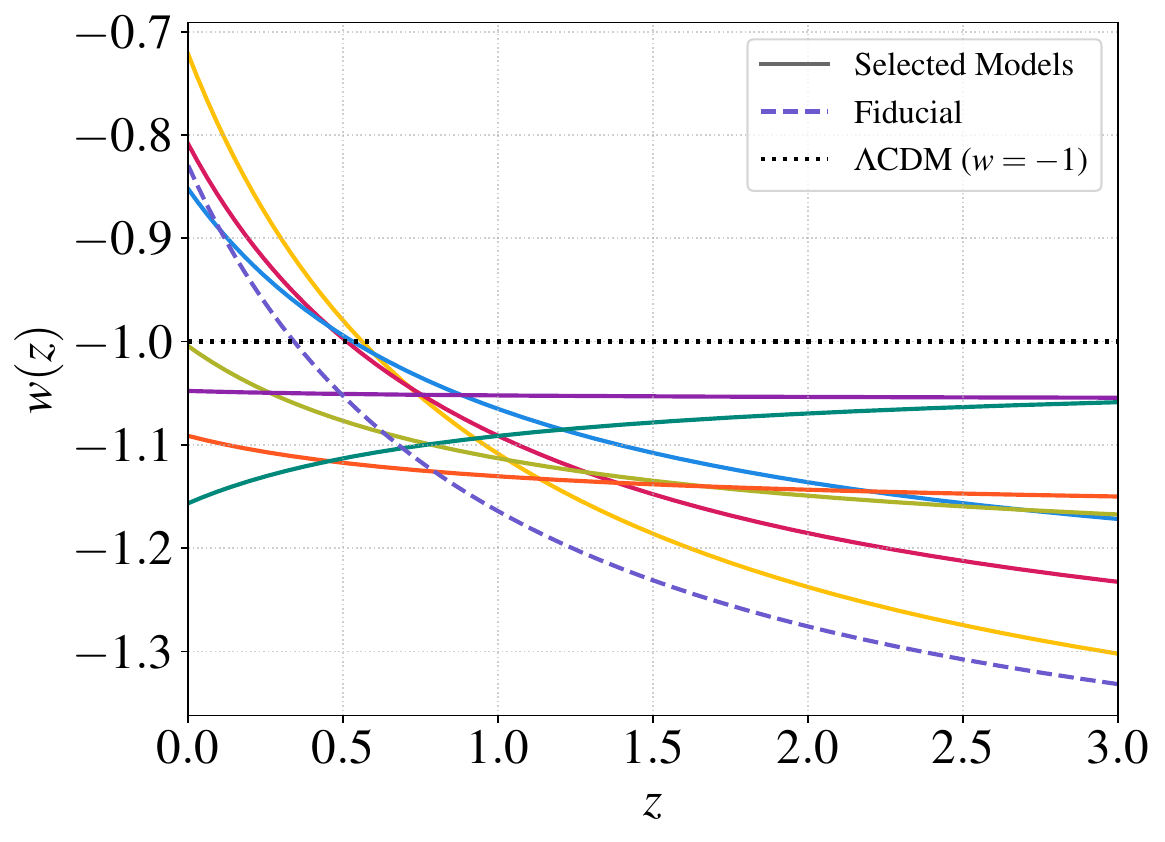}
    \caption{Equation of state of dark energy as a function of redshift for the models plotted in Fig.~\ref{fig:jhat4bins}. The colors refer to the same models as in that figure.}
    \label{fig:w_z}
\end{figure}

\begin{figure}
    \centering
    \includegraphics[width=1\linewidth]{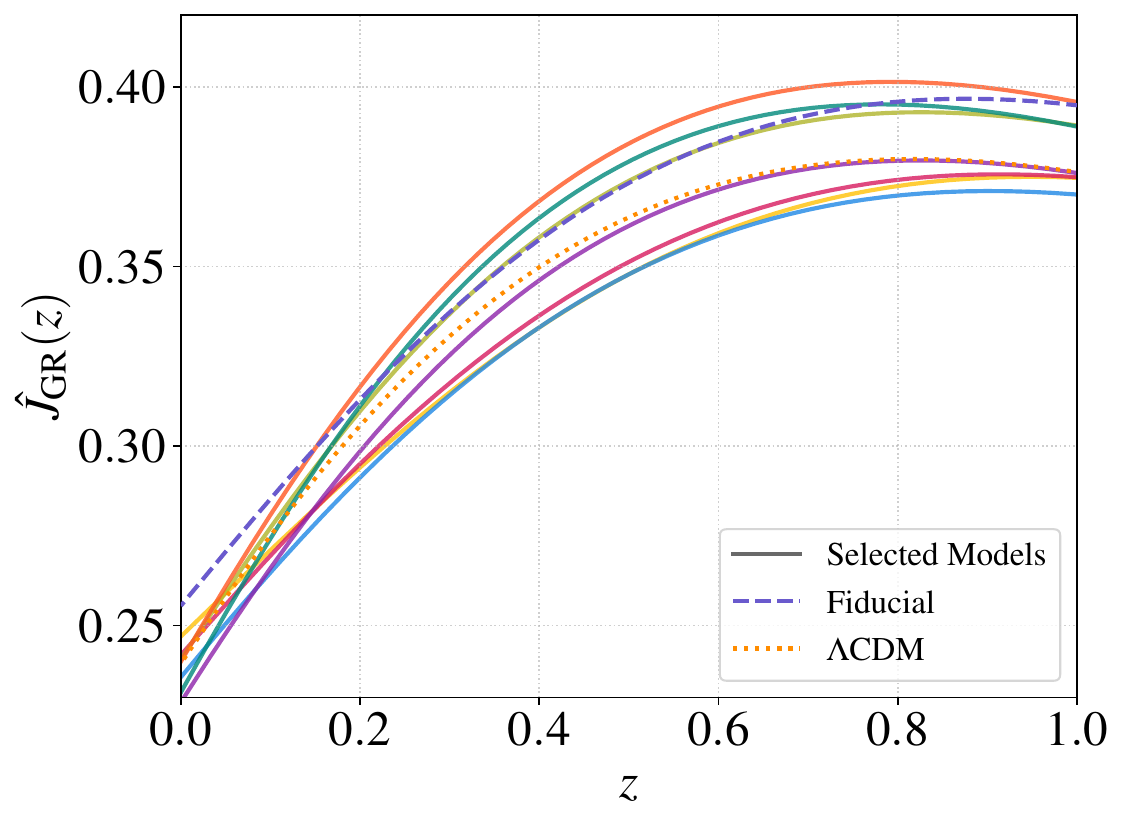}
    \caption{GR prediction for the Weyl potential, $\hJ_{\rm GR}$, as a function of redshift for the models plotted in Fig.~\ref{fig:jhat4bins}. The colors refer to the same models as in that figure.} 
    \label{fig:JGR_z}
\end{figure}

\begin{figure}
    \centering
    \includegraphics[width=0.95\linewidth]{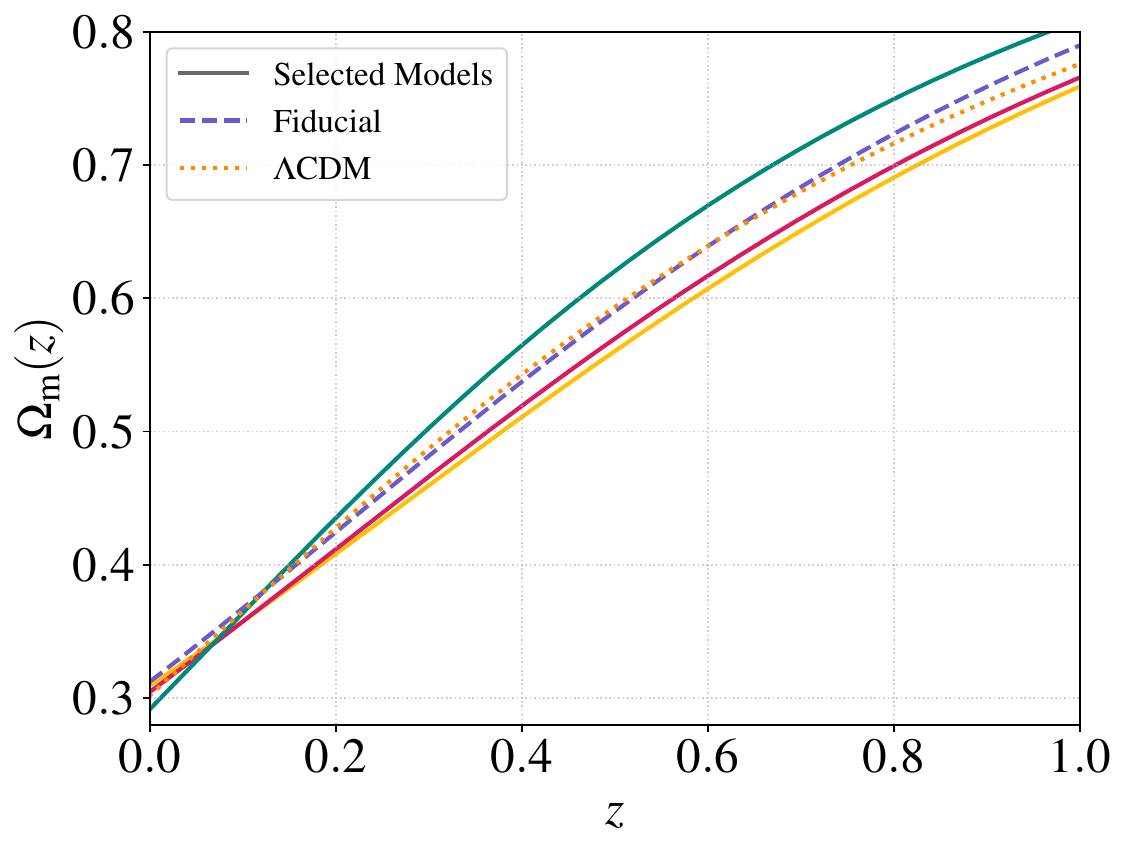}
    \caption{Evolution of $\Omega_{\rm{m}}(z)$ for Models 1, 2 and 7 represented with the same colors as in Fig.~\ref{fig:jhat4bins}. Here we removed some of the models plotted in Fig.~\ref{fig:jhat4bins} to improve visual clarity.}
    \label{fig:Om_z}
\end{figure}

In Fig.~\ref{fig:chi2} we show the chi-square in the second redshift bin for models on the surface $\{\Omega_{\rm m}, w_0, w_a\}^{\rm DESI}$. We have discretized $w_0$ and $w_a$ in bins of size (0.02, 0.05), respectively. Fig.~\ref{fig:w_z} shows the equation of state $w(z)$ for a selection of models that solve the tension in the second bin, while Fig.~\ref{fig:JGR_z} shows the GR prediction $\hJ_{\rm GR}$ for these models. Finally, Fig.~\ref{fig:Om_z} shows the matter density parameter evolution $\Omega_{\rm{m}} (z)$ for some particular models.

\clearpage

\bibliography{Jhat_w0wa}
\bibliographystyle{apsrev4-1}

\end{document}

%% file: camille_setup.tex
\usepackage{ifthen} 
\usepackage{fp,calc}
\usepackage{lastpage}
\IfPackageLoadedTF{color}{}{\usepackage{color}}

\usepackage{fancyhdr}
\usepackage{adjustbox}
\usepackage[english]{babel}

\IfPackageLoadedTF{xcolor}{}{\usepackage[svgnames, table]{xcolor}}
\usepackage{layout}
\usepackage{enumitem}
\usepackage[normalem]{ulem}
\usepackage{parskip}
\usepackage{multirow}
\usepackage{comment}
\usepackage{booktabs}
\usepackage[flushleft]{threeparttablex}

\newboolean{articletitles}
\setboolean{articletitles}{true} 
\newboolean{allauthors}
\setboolean{allauthors}{false} 

\providecommand\prd{Phys.\ Rev.\ D}

\providecommand\jcap{JCAP }
